\documentclass[twocolumn,superscriptaddress, floatfix, amsfonts, aps]{revtex4-1} 
\usepackage{amsmath}
\usepackage{bm}
\usepackage{graphicx}
\usepackage{subfigure}
\usepackage{float}
\usepackage{mathrsfs}
\usepackage{footmisc}
\usepackage{multirow}
\usepackage{graphicx}
\usepackage{booktabs, ctable}
\usepackage{soul,xcolor}
\usepackage{xcolor, ulem, color, framed}
\usepackage{amssymb}
\usepackage[colorlinks, linkcolor=red,anchorcolor=green,citecolor=blue]{hyperref}

\usepackage[colorlinks, linkcolor=red,anchorcolor=green,citecolor=blue]{hyperref}

\graphicspath{{./figs/}}

\begin{document}

 \title{Non-thermal distributions of charm and charmonium in relativistic heavy-ion collisions}

  \author{Chaoyu Pan}
 \author{Shuhan Zheng}
  \author{Meimei Yang}
 \author{Zhiwei Liu}
 \author{Baoyi Chen}\email{baoyi.chen@tju.edu.cn}
\affiliation{Department of Physics, Tianjin University, Tianjin 300354, China}

\begin{abstract}
We employ the Boltzmann transport model to study the charmonium regeneration with non-thermal charm quarks in relativistic heavy-ion collisions. As heavy quarks do not reach kinetic thermalization in the quark-gluon plasma (QGP), the final transverse momentum distribution of regenerated charmonium depends on the degree of charm quark kinetic thermalization. When the charm momentum distribution becomes harder, more charm quarks are distributed in the middle and high $p_T$, where the production of regenerated charmonium also becomes larger. With non-thermal momentum distribution of charm quarks in their coalescence process, the nuclear modification factor $R_{AA}(p_T)$ of charmonium enhances at middle $p_T$. 
Besides, the elliptic flow $v_2(p_T)$ of charmonium also enhances at middle $p_T$ as more regenerated charmonium are distributed in this $p_T$ region. The theoretical calculations with non-thermal charm distribution explain well the $p_T$ dependence of charmonium $R_{AA}$ and $v_2$, which indicates that charm quarks do not reach complete kinetic thermalization in the QGP when charmonium are regenerated.

\end{abstract}
\date{\today}

 \maketitle
 
%=======================================

\section{introduction}
In relativistic heavy-ion collisions, it is believed that an extremely hot and deconfined medium made up of quarks and gluons, called ``Quark-Gluon Plasma'' (QGP) is produced~\cite{Bazavov:2011nk}. The study of signals~\cite{Matsui:1986dk,Stoecker:2004qu} and properties~\cite{Ding:2015ona} of this new deconfined matter, such as transport coefficients and initial energy densities helps to comprehend the strong interaction at finite temperatures. 
Heavy quarks and quarkonium are predominantly produced in the nuclear parton hard scatterings due to their large masses. They have been proposed as clean probes of early stage of the hot medium generated in nucleus-nucleus (AA) collisions in the Large Hadron Collider (LHC) and the Relativistic Heavy Ion Collider (RHIC)~\cite{Zhao:2020jqu,Zhao:2011cv,He:2012df,Yan:2006ve,Liu:2010ej,Chen:2019qzx}.  In this hot medium, the heavy quark potential is screened by thermal light partons~\cite{Wen:2022utn,Wen:2022yjx}, which results in the ``melt'' of quarkonium-bound states when the medium temperature is sufficiently high~\cite{Karsch:2005nk,Satz:2005hx}. The highest temperature at which quarkonium-bound states can survive is called the ``dissociation temperature'' $T_d$~\cite{Liu:2012zw,Guo:2012hx,Chen:2013wmr}. Below this $T_d$, inelastic collisions from 
thermal light partons can also dissociate the quarkonium-bound states~\cite{Adil:2006ra,Wong:2001td,Zhu:2004nw}. The survival probability of quarkonium decreases when traveling through the QGP due to both color screening and parton inelastic scatterings. The hot medium effects are characterized by the nuclear modification factor $R_{AA}$, which is defined as the ratio of quarkonium production in AA collisions to the production in proton-proton (pp) collisions scaled by the number of nucleon binary collisions $N_{coll}$. 

At the LHC, where there are many heavy quarks in the QGP, charm and anti-charm quarks have a significant chance to combine into new bound states when they move to areas with lower temperatures. 
This process becomes even more dominant in the production of charmonium with the presence of more charm pairs~\cite{Du:2017qkv,Zhao:2017yan}. 
As charm quarks lose energy as they travel through the QGP~\cite{Braaten:1991we,Gyulassy:2000fs,Djordjevic:2003zk,Qin:2007rn}, the $p_T$ spectrum of regenerated charmonium depends on the momentum distribution of charm quarks before their coalescence into charmonium states. When charm quarks have a non-thermal distribution, the mean $p_T$ of regenerated charmonium will be larger than that of the thermal case.

This work studies the $p_T$ dependence of charmonium $R_{AA}(p_T)$ and $v_2(p_T)$ by employing various momentum distributions of charm quarks. Cold nuclear matter effects have been included in the initial distributions of charm quarks and charmonium. Hot medium evolution is described using  hydrodynamic equations. 
The manuscript is organized as follows. In Section II, the transport model is introduced. In Section III, non-thermal distributions of charm quarks are discussed. Hot medium evolution is given as a background in Section IV. 
 In Section V, the scaled $R_{AA}$ and $v_2$ of charmonium are calculated and compared with experimental data for different non-thermal momentum distributions of charm quarks. Lastly, a final summary is given in Section VI.

\section{transport model}

The distribution of heavy quarkonium in a medium has been extensively studied using various models, including transport models~\cite{Grandchamp:2003uw,Du:2015wha,Chen:2016dke,Yao:2020eqy,Yao:2018sgn}, coalescence models~\cite{Zhao:2017yan, Chen:2021akx, Thews:2000rj}, statistical Hadronization model~\cite{Braun-Munzinger:2000csl}, open quantum system approaches~\cite{Yao:2021lus,Brambilla:2020qwo, Akamatsu:2018xim}, etc. Transport model takes into account both primordial production and regeneration. The Boltzmann transport equation for charmonium evolution is written as~\cite{Yan:2006ve}, 
\begin{align}
    \partial_t f_\psi +{\bf v}\cdot {\bf \bigtriangledown_x} f_\psi 
     = -\alpha f_\psi +\beta,
\end{align}
where the distribution of charmonium in phase space, denoted by $f_\psi$, varies over time due to charmonium diffusion which is represented by the term ${\bf v}\cdot {\bf \bigtriangledown_x} f_\psi$. ${\bf v}$ is the velocity of charmonium. The decay rate $\alpha$ accounts for inelastic collisions and the color screening effect that can dissociate charmonium states. This rate depends on both the density of light partons and the inelastic cross sections~\cite{Zhu:2004nw}, 
\begin{align}
\label{alphabeta}
\alpha({\bm p},x) &= {1\over 2E_T}\int{d^3{\bm p}_g\over(2\pi)^32E_g}W_{g\psi}^{c\bar c}(s)f_g({\bm p}_g,x) \Theta(T(x)-T_c),
\end{align}
where ${\bf p}_g$ and $E_g$ are the momentum and the energy of thermal gluons. $E_T=\sqrt{m_\psi +p_T^2}$ is the transverse energy of charmonium with the mass $m_{J/\psi}=3.1$ GeV. $f_g$ is the Bose distribution of massless gluons. The step function $\Theta(T-T_c)$ ensures that 
the gluon-dissociation process only happens above the critical temperature $T_c$ of the deconfined phase transition. 
$W_{g\psi}^{c\bar c}=4\sigma_{g\psi}^{c\bar c}(s)F_{g\psi}(s)$ is the charmonium dissociation probability in the reaction $g+J/\psi\rightarrow c+\bar c$, where $s$ is the center-of-mass energy of the gluon and the charmonium. $F_{g\psi}$ is the flux factor. $\sigma_{g\psi}^{c\bar c}$ is the gluon-dissociation cross section. 
It is obtained via the operator-production-expansion (OPE) method~\cite{Peskin:1979va,Bhanot:1979vb}, 
\begin{eqnarray}
\sigma_{g-J/\psi}(\omega)&=&A_0 {(x-1)^{3/2}\over x^5}.
\end{eqnarray}
We use $w$ to represent the gluon energy, while $x$ refers to the ratio of the gluon energy to the $J/\psi$ in-medium binding energy $\epsilon_\psi$. The constant factor 
$A_0=2^{11}\pi/(27\sqrt{m_c^3\epsilon_\psi})$ has charm quark mass $m_c=1.87$ GeV. For loosely bound excited states such as $\chi_c(1P)$ and $ \psi^\prime$, we obtain the decay widths using the geometry scale~\cite{Chen:2018kfo}.

Above the critical temperature, heavy quark potential is partially restored~\cite{Lafferty:2019jpr,Liu:2018syc}, allowing charm and anti-charm quarks to combine and form new bound states through the reaction $c+\bar c\rightarrow J/\psi +g$. The regeneration rate of charmonium $\beta$ is proportional to the densities of charm and anti-charm quarks, 
\begin{align}
\label{lab-beta}
\beta({\bm p},x) &= {1\over 2E_T}\int {d^3{\bm p}_g\over(2\pi)^32E_g}{d^3{\bm p}_c \over(2\pi)^32E_c}{d^3{\bm p}_{\bar c}\over(2\pi)^32E_{\bar c}} \nonumber \\
&\quad\times W_{c\bar c}^{g\psi}(s) f_c({\bm p}_c,{\bf x})f_{\bar c}({\bm p}_{\bar c},{\bf x})\nonumber\\
&\quad \times \Theta(T({\bf x})-T_c)(2\pi)^4\delta(p+p_g-p_c-p_{\bar c}),
\end{align}
where ${\bf p}_c$ and ${\bf p}_{\bar c}$ represent the momentum of charm and anti-charm quarks respectively. $E_c=\sqrt{m_c^2 + p_c^2}$ denotes the energy of charm quark. The probability of combination of $c$ and $\bar c$, denoted by $W_{c\bar c}^{g\psi}$, is determined through the detailed balance. The delta function $\delta(p+p_g-p_c-p_{\bar c})$ ensures a four energy-momentum conservation in the reaction.  
Charm quark distribution $f_c({\bf p}_c, {\bf x})$ in the QGP will be obtained by fitting the results from the Langevin model~\cite{He:2014cla,Qin:2010pf,Xu:2017obm}.

\section{non-thermal distribution of charm quarks}

Heavy quarks experience significant energy loss when they move through a hot medium. 
The Langevin equation can be used to obtain the spatial and momentum distributions of heavy quarks. 
For simplicity, we assume that the heavy quark distribution in phase space can be separated into a production of the spatial density $\rho_c({\bf x},t)$ and the momentum distribution as $f_c({\bf p}_c, {\bf x},t)=\rho_c({\bf x},t) f({\bf p}_c, t)$. The spatial density $\rho_c({\bf x}, t)$ decreases with the times in the expanding QGP. It mainly affects the yield of regenerated charmonium.  
Therefore, we approximate the spatial density with the diffusion equation, $\partial_\mu (\rho_c u^\mu)=0$~\cite{Zhou:2014kka}, which is usually used in the limit of the kinetic thermalization.  The heavy quark spatial density only depends on the four-velocity of the 
medium, which is given by hydrodynamic equations. 
The initial profile of $\rho_c$ is proportional to the 
$d\sigma_{pp}^{c\bar c}/dy T_A({\bf x}_T-{\bf b}/2)T_B({\bf x}_T+{\bf b}/2)$, where $d\sigma_{pp}^{c\bar c}/dy=(1.165,\ 0.718)\ \rm{mb}$~\cite{ALICE:2021dhb,LHCb:2016ikn} are the rapidity differential 
cross sections of charm quarks in central and forward rapidities respectively. $T_{A(B)}$ is the nuclear thickness function.

\begin{figure}[!htb]
\includegraphics[width=0.45\textwidth]{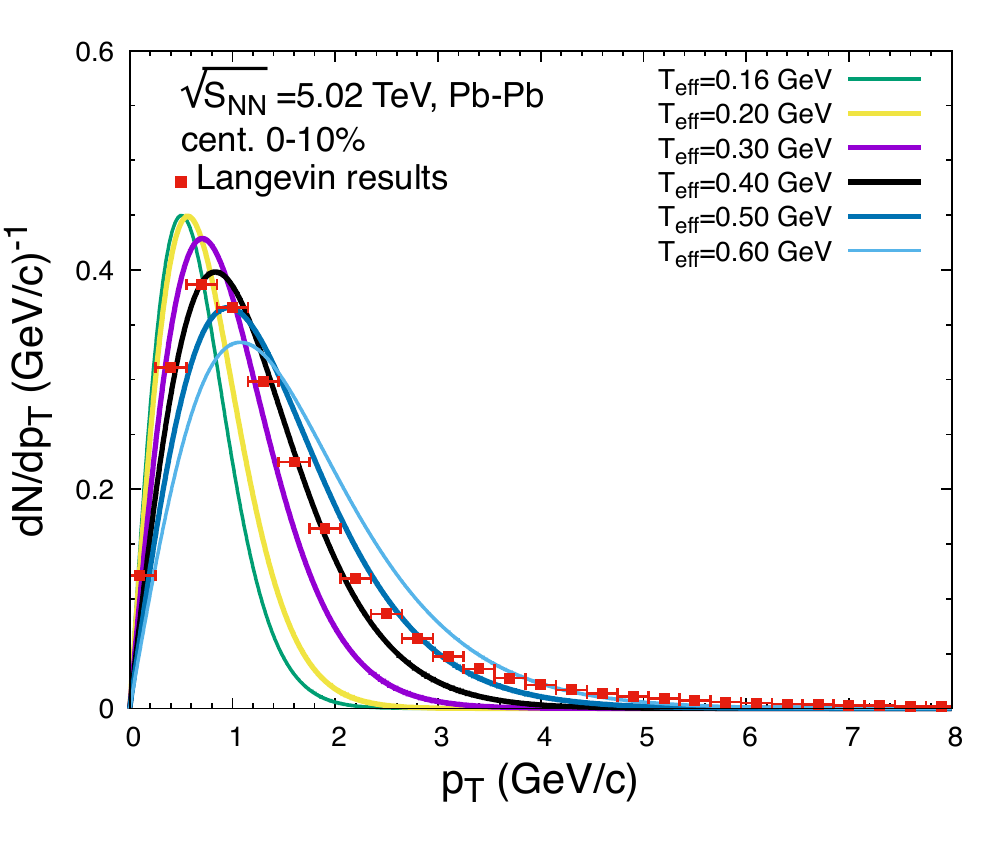}
\caption{(Color online) Normalized transverse momentum distribution $dN/dp_T$ of charm quarks in the most central Pb-Pb collisions at  $\sqrt{s_{NN}}=5.02$ TeV. Dots are the normalized $p_T$ distribution of charm quarks given by the Langevin equation on the hypersurface specified with the $J/\psi$ regeneration temperature $T_{\rm rege}({\bf x})\sim 1.2\ T_c$. The spatial diffusion coefficient is taken to be $\mathcal{D}_s(2\pi T)=5$. Different Fermi distributions are plotted where $T_{\rm eff}$ controls the charm quark momentum distribution. 
}
\label{lab-c-fist}
\end{figure}

As for the momentum distribution, charm quarks do not reach kinetic thermalization during charmonium regeneration. The realistic momentum distribution of charm quarks in the QGP is obtained by the event-by-event simulations of the Langevin equation~\cite{Cao:2015hia,He:2019vgs}. Charm quarks are randomly regenerated according to the momentum distribution obtained from FONLL calculation~\cite{Cacciari:1998it,Cacciari:2001td}. These charm quarks are evolved in QGP with the Langevin equation, where heavy quark energy loss is determined by the spatial diffusion coefficient $\mathcal{D}_s(2\pi T)=5$, which has been previously used for open heavy flavor hadrons~\cite{He:2012df,Rapp:2018qla,Dong:2019unq}. When charm quarks move to regions with a relatively low temperature, the heavy quark potential is partially restored and $c$ and $\bar c$ can combine into a bound state. Most $J/\psi$ are regenerated below the temperature $T_{\rm rege}\sim 1.2\ T_c$ where the heavy quark potential is mostly restored at the distance of the $J/\psi$ radius~\cite{Satz:2005hx,Chen:2018kfo}. The normalized momentum distribution of charm quarks on the hypersurface with $T({\bf x})=T_{\rm rege}$ is given by the Langevin equation. It is plotted as dots in Fig.\ref{lab-c-fist}.

To incorporate the non-thermal distribution of charm quarks into the transport model, we use the Fermi distribution to simulate a distribution that is close to the realistic distribution given by the Langevin model. 
The $T_{\rm eff}$ parameter in the Fermi distribution can be determined based on the realistic distribution of charm, 
\begin{align}
    \label{lab-fc}
    f_c({\bf p}_c)
    = {N_{\rm norm}\over e^{u\cdot p_c/T_{\rm eff}}+1},
\end{align}
where $T_{\rm  eff}$ does not represent the temperature of the medium anymore. It characterizes the realistic momentum distribution of charm quarks in the regeneration process and is connected with the charm energy loss in the medium. $N_{\rm norm}$ is the normalization factor satisfying the relation $\int d{\bf p}_c f_c({\bf p}_c)=1$. $u$ and $p_c$ are the four-velocity and four-momentum of the fluid and charm quarks respectively. 
We take the spatial diffusion coefficient 
$\mathcal{D}_s(2\pi T)=5$ to evolve charm quarks in the QGP and stop the evolution at $T({\bf x})=T_{\rm rege}$.
In Fig.\ref{lab-c-fist}, the normalized final transverse momentum distribution of charm quarks in the most central Pb-Pb collisions is plotted with dots. 
Different Fermi distributions are also plotted where $T_{\rm eff}$ is taken as different values. When $T_{\rm eff}$ is larger than the medium temperature $T_{\rm rege}$, it indicates that 
charm quarks are further away from the kinetic thermalization. The line with $T_{\rm eff}\sim 0.4$ GeV fits the Langevin results better than other lines. Furthermore, the degree of charm kinetic thermalization differs in different collision centralities. We will take different transverse momentum distributions of charm quarks in the charmonium regeneration.

\section{hot medium evolution}

Relativistic heavy ion collisions produce a hot, deconfined medium that behaves like a nearly perfect fluid. The dynamical evolution of the medium can be described with hydrodynamic equations. We assume that the longitudinal expansion of the hot medium follows a Bjorken evolution and use 2+1 dimensional ideal hydrodynamic equations to simulate its expansion, 
\begin{align}
\partial_{\mu\nu} T^{\mu\nu}=0.
\end{align}
The energy-momentum tensor is $T^{\mu\nu}=(e+p)u^\mu u^\nu -g^{\mu\nu}p$. $u^\mu$ is the four-velocity of the medium. Energy density $e$ and the pressure $p$ change with the times and the position. An equation of the state of the medium is necessary to close the equations. QGP is treated as an ideal gas made up of massless $u,d$ quarks, gluons and the strange quark with a mass of $m_s=150$ MeV. Meanwhile the hadronic medium is treated as an ideal gas composed of all known hadrons and resonances with mass up to 2 GeV~\cite{ParticleDataGroup:2020ssz}. The phase transition between the QGP and the hadronic gas is a first-order phase transition with a critical temperature $T_c=165$ MeV~\cite{Sollfrank:1996hd}. The initial energy density of the medium is determined by the final charged multiplicity, where the maximum initial temperature at the center of the fireball is extracted to be $T_0(\tau_0, {\bf x}_T=0|{\bf b}=0)=510$ MeV at the starting time of hydrodynamic equations in the most central $5.02$ TeV Pb-Pb collisions~\cite{Zhao:2017yhj}. $\tau_0=0.6$ fm is the time scale of the medium reaching local equilibrium~\cite{Shen:2014vra}. The initial temperatures at other positions ${\bf x}_T$ are obtained with the Glauber model.

\section{Initial conditions}
At the beginning of nuclear collisions, charmonium production is produced in parton hard scatterings. Their production can be treated as the production in pp collisions scaled with the number of binary collisions $N_{coll}$. 
In pp collisions without hot medium effects, the momentum distribution of $J/\psi$ has been measured in experiments, which can be parametrized using a power law function, 
\begin{align}
    {d^2\sigma_{pp}^{J/\psi}\over dy 2\pi p_T dp_T}={(n-1)\over \pi (n-2)\langle p_T^2\rangle_{pp}}[1+{p_T^2\over (n-2)\langle p_T^2\rangle_{pp}}]^{-n}{d\sigma_{pp}^{J/\psi}\over dy},
\end{align}
where the parameters are fitted to be 
$\langle p_T^2\rangle_{pp}=12.5\ \rm{(GeV/c)^2}$ and $n=3.2$ in the central rapidities, and the 
rapidity differential cross section of inclusive $J/\psi$ is taken to be $d\sigma_{pp}^{J/\psi}/dy=5.2\ \mu b$ (in central rapidity)~\cite{ALICE:2012vup,ALICE:2012vpz,CMS:2018gbb}. In the forward rapidity, the differential cross section is fitted to be $d\sigma_{pp}^{J/\psi}/dy=3.25\ \mu b$. For excited states of charmonium, their normalized $p_T$ distribution is the same as the distribution of $J/\psi$ due to the small difference between their masses. The inclusive charmonium consists of both the non-prompt part from B decay, and the prompt part which is defined as the sum of the primordial production and the regeneration.  The inclusive nuclear modification factor related to the prompt and the non-prompt $R_{AA}$ is as follows~\cite{Chen:2013wmr}, 
\begin{align}
    R_{AA}^{incl}={R_{AA}^{prompt}\over 1+r_B} + {R_{AA}^{B}r_B\over 1+r_B},
\end{align}
where the $R_{AA}^{prompt}$ and $R_{AA}^B$ is the prompt and the non-prompt nuclear modification factors. $r_B=f_B/(1-f_B)$ is the ratio of non-prompt and prompt charmonium production in pp collisions. The fraction of non-prompt $J/\psi$ in the inclusive production is fitted to be $f_B=0.04+0.23(p_T/\rm{(GeV/c)})$~\cite{CDF:2004jtw,CMS:2010nis,ALICE:2012vpz}, without clear dependence on the collision energy. 
$R_{AA}^{prompt}$ is calculated with the transport model which includes charmonium dissociation and regeneration. While in the non-prompt part, the final distribution of non-prompt $J/\psi$ is connected with the bottom quark energy loss in the hot medium. The non-prompt nuclear modification factor $R_{AA}^B$ can be extracted from the Langevin equation~\cite{Yang:2023rgb}. 

Charmonium initial distribution in AA collisions is also modified by the cold nuclear matter effects compared with the case in pp collisions. The partons scatter with other nucleons to obtain extra energy before fusing into charm pairs and charmonium. This process is called the Cronin effect~\cite{Cronin:1974zm}, making the momentum distribution of primordial charmonium become harder than the distribution extracted in pp collisions. This is included via the modification $\langle p_T^2\rangle_{pp}+a_{gN}\langle l\rangle$. Here $\langle l\rangle $ is the average path length of partons traveling through the nucleus before the hard scattering. The transverse momentum square parton obtained per unit length is denoted by $a_{gN}$. The value of $a_{gN}$ is determined to be $a_{gN}=0.15\ \rm{(GeV/c)^2}$~\cite{Chen:2016dke,Zhou:2014kka} based on the charmonium momentum broadening observed in proton-nucleus collisions.  Another important cold nuclear matter effect is the shadowing effect~\cite{Mueller:1985wy}. Parton densities in the nucleons of the nucleus become different compared to those in free nucleons. This effect results in the suppression of parton density and hence the reduction of the number of charmonium and charm pairs in heavy-ion collisions. The shadowing effect is calculated to be 0.6 in the most central Pb-Pb collisions at $\sqrt{s_{NN}}=5.02$ TeV with the EPS09 package~\cite{Eskola:2009uj}. The shadowing factor at other centralities can be obtained via the scale of the thickness function~\cite{Zhou:2014kka}.

\section{charmonium distribution in heavy-ion collisions}
In the regeneration, when the charm quark momentum distribution becomes different, 
the $p_T$ distribution of regenerated charmonium from the reaction $c+\bar c\rightarrow J/\psi +g$ also becomes different. Additionally, the total production of charmonium depends on the degree of charm kinetic thermalization. To focus on the shape of charmonium $p_T$ distribution with different non-thermal distributions of charm quarks, the total yields of regeneration are scaled to the same value to fit the experimental data of $R_{AA}(N_p)$ in the most central collisions. With this disposure, we can clearly see how the $p_T$ distribution of charmonium $R_{AA}(p_T)$ and $v_2(p_T)$ are affected by the non-thermal charm quarks, especially at middle and high $p_T$. 

\begin{figure}[!htb]
\includegraphics[width=0.42\textwidth]{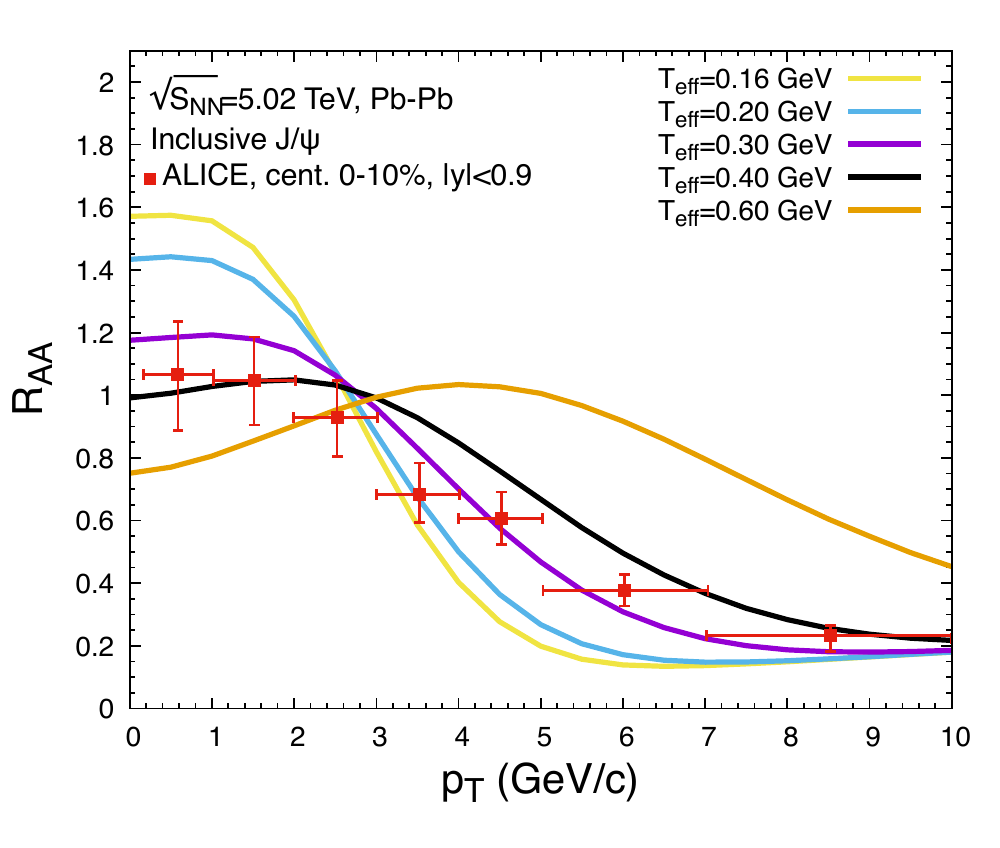}
\caption{(Color online) $p_T$ dependence of inclusive $J/\psi$ nuclear modification factor in the central rapidity of $\sqrt{s_{NN}}=5.02$ TeV Pb-Pb collisions. Different non-thermal distributions of charm quarks are considered by taking $T_{eff}=(0.16, 0.2, 0.3,0.4,0.6)$ GeV in the collision centrality 0-10\% where regeneration contribution becomes dominant. The primordial production, regeneration, and decay from B-hadron have been included in the lines. 
The experimental data is cited from the ALICE Collaboration~\cite{ALICE:2019lga}. 
}
\label{lab-RAA-pt-2}
\end{figure}

In Fig.\ref{lab-RAA-pt-2}, we calculated the $J/\psi$ nuclear modification factor as a function of $p_T$ in the centrality 0-10\% in $\sqrt{s_{NN}}=5.02$ TeV Pb-Pb collisions, where regeneration dominates the total production of $J/\psi$. With smaller $T_{\rm eff}$, charm quark momentum distribution before regeneration becomes softer, leading to regenerated charmonia carry small $p_T$ and are distributed in low $p_T$ region. At middle and high $p_T$, regeneration contribution becomes negligible as the density of charm quarks is small, shown as the lines labeled with $T_{\rm eff}=0.2$ and $0.16$ GeV. 
These lines with softer charm distributions underestimate the experimental data at middle $p_T$. When the momentum distribution of charm quarks becomes harder, such as the cases with $T_{\rm eff}=(0.3,0.4)$ GeV, regenerated $J/\psi$ tends to carry larger $p_T$ and enhance the inclusive production of $J/\psi$ at $p_T\sim 4-5$ GeV/c, shown as the lines $T_{\rm eff}=(0.3,0.4)$ GeV. However, when the charm momentum distribution used in the regeneration becomes extremely hard ($T_{\rm eff}=0.6$ GeV), the $J/\psi$ inclusive $R_{AA}$ enhances significantly at high $p_T$ which is far away from the realistic case. 
Inspired by the realistic distributions of charm quarks given by the Langevin equation in Fig.\ref{lab-c-fist}, the value of $T_{\rm eff}$ is expected to be around $T_{\rm eff}\sim 0.4$ GeV. We can see that the shape of the line with $T_{\rm eff}=0.3\sim 0.4$ GeV can give a better explanation of the data compared to the case of completely kinetic thermalization ($T_{\rm eff}=0.16\sim 0.2$ GeV). 

\begin{figure}[!htb]
\includegraphics[width=0.42\textwidth]{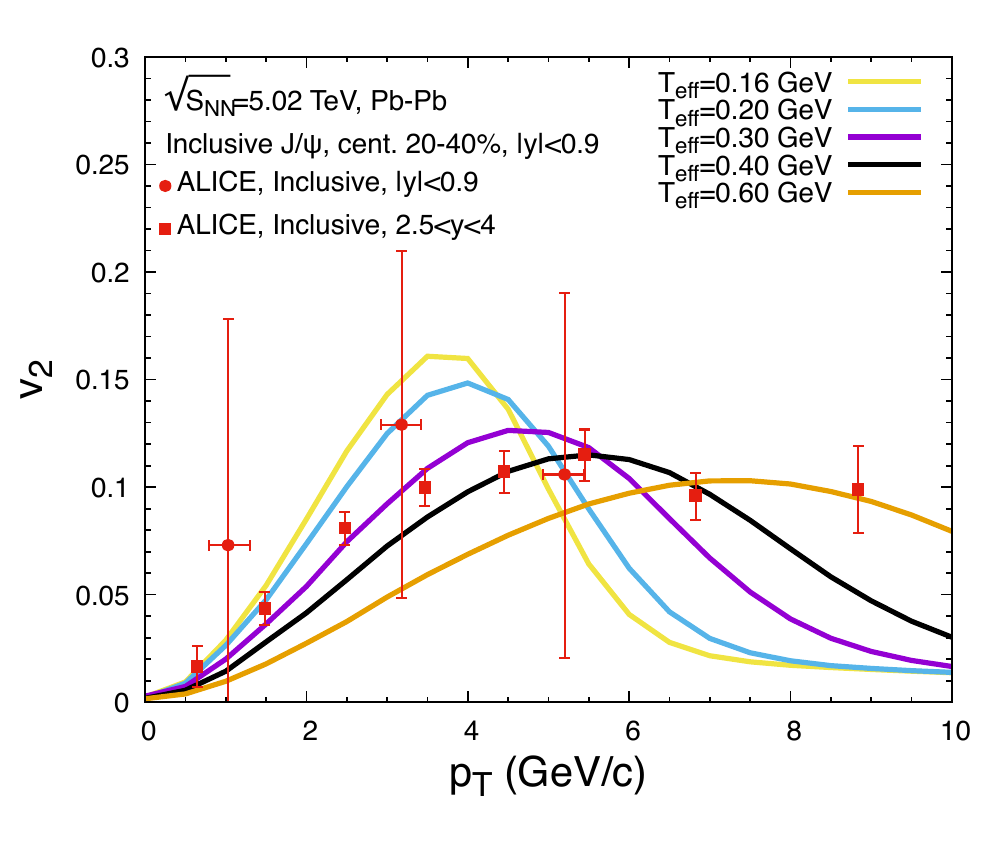}
\caption{(Color online) $p_T$ dependence of inclusive $J/\psi$ elliptic flows in the forward rapidity of $\sqrt{s_{NN}}=5.02$ TeV Pb-Pb collisions. Different non-thermal distributions of charm quarks are considered by taking $T_{eff}=(0.16, 0.2, 0.3,0.4,0.6)$ GeV. 
The experimental data is cited from ALICE Collaboration~\cite{ALICE:2017quq,ALICE:2020pvw}.  }
\label{lab-v2-p}
\end{figure}

In Fig.\ref{lab-v2-p}, the elliptic flows of inclusive $J/\psi$ are also calculated by taking different non-thermal distributions of charm quarks. 
Primordially produced charmonium carries small elliptic flows when moving along different trajectories in the anisotropic medium. Charmonium dissociation along different paths results in a non-zero $v_2$ through medium dissociation. This effect becomes important at large $p_T$ where primordial production dominates the total yield of $J/\psi$. 
On the other hand, charm quarks are strongly coupled with QGP. Charm quarks develop collective flows via elastic scatterings with thermal partons, which will be inherited by the regenerated $J/\psi$. 
Therefore, when the charm momentum distribution becomes harder, the regenerated $J/\psi$s are mainly  distributed at higher $p_T$. The elliptic flows of inclusive $J/\psi$ are enhanced by the regeneration. When charm quarks are kinetically thermalized, they satisfy a normalized Fermi distribution with the medium temperature $T_{\rm rege}$. In the non-thermal cases, charm distributions become hard, and are characterized by a parameter $T_{\rm eff}$ which is usually larger than the medium temperature. Lines with $T_{\rm eff}\sim 0.4$ GeV can better explain the large $v_2$ at $p_T\sim 6$ GeV/c than cases with softer charm distributions. This indicates that regeneration is still important in this middle $p_T$ region.

From the shapes of $R_{AA}(p_T)$ and $v_2(p_T)$, one can see that regeneration with non-thermal momentum distributions of charm quarks becomes more important at middle and high $p_T$. 
This will increase the elliptic flows of $J/\psi$ at $p_T\sim 6$ GeV/c. 
While in the thermal case, the contribution of regeneration decreases rapidly with increasing $p_T$. The parameter $T_{\rm eff}$ characterizing the degree of charm kinetic thermalization is 
extracted to be around $T_{\rm eff}= 0.3\sim 0.4$ GeV, which is larger than the medium temperatures of charmonium regeneration. Charm momentum distribution is expected to be non-thermal when charmonium regeneration happens, even when $D$ mesons are close to kinetic thermalization after experiencing the whole evolution of the hot medium.

\section{Summary}

We employ the transport model to study the $p_T$ distribution of charmonium with non-thermal charm quarks in $\sqrt{s_{NN}}=5.02$ TeV Pb-Pb collisions. As the heavy quark potential is partially restored above the critical temperature $T_c$, charmonium can be regenerated above $T_c$ where charm quarks do not reach complete kinetic thermalization. The realistic momentum distribution of charm quarks in charmonium regeneration is simulated with the Langevin equation. To incorporate non-thermal charm distributions into the transport model, we introduce a parameter $T_{\rm eff}$ in a normalized Fermi distribution, where $T_{eff}$ is determined by fitting the realistic  distribution of charm quarks from the Langevin model. The value of $T_{\rm eff}$ is found to be larger than the temperature of $J/\psi$ regeneration, which indicates that the charm quark distribution is harder than that in the thermal case. We take different values of $T_{\rm eff}$ in the transport model to calculate $R_{AA}(p_T)$ and $v_2(p_T)$. The regeneration of charmonium with different non-thermal distributions has been scaled to fit the $R_{AA}(N_p)$ of $J/\psi$, which allows us to focus on the shapes of $R_{AA}(p_T)$ and $v_2(p_T)$ when taking different momentum distributions of charm quarks. 
The $R_{AA}$ and $v_2$ enhance evidently at middle $p_T$ when taking a non-thermal charm distribution, as more charm quarks and regenerated charmonium are distributed in middle and high $p_T$ region. This helps to extract the realistic momentum distribution of charm quarks in charmonium regeneration.

\vspace{1cm}
\noindent {\bf Acknowledgement}: Chaoyu Pan and Shuhan Zheng contributed equally to this work. This work is supported by the National Natural Science Foundation of China
(NSFC) under Grant Nos. 12175165.

\end{document}